\documentclass[prl,twocolumn,floatfix,aps]{revtex4}
\usepackage{amsmath}
\usepackage{makeidx}
\usepackage{amssymb}
\usepackage{graphicx}

\usepackage[usenames]{color}
\usepackage[normalem]{ulem}

\usepackage{hyperref}
\usepackage[T1]{fontenc} 

\begin{document}

\title{Supersolid-like states in a two-dimensional trapped spin-orbit-coupled spin-1 condensate}


\author{S. K. Adhikari\footnote{s.k.adhikari@unesp.br      \\  https://professores.ift.unesp.br/sk.adhikari/}}
\affiliation{Instituto de F\'{\i}sica Te\'orica, Universidade Estadual Paulista - UNESP, 01.140-070 S\~ao Paulo, S\~ao Paulo, Brazil}
      

\date{\today}

\begin{abstract}

We study supersolid-like states  in a quasi-two-dimensional trapped Rashba and Dresselhaus spin-orbit (SO) coupled spin-1 condensate.    For small strengths of SO coupling $\gamma$ ($\gamma \lessapprox 0.75$), in the ferromagnetic phase, circularly-symmetric $(0,\pm 1, \pm 2)$- and $(\mp 1, 0,\pm 1)$-type states are formed where the numbers in  the parentheses denote the angular momentum of the vortex at the center of the components and where the upper (lower) sign correspond to Rashba (Dresselhaus) coupling; in the  antiferromagnetic phase,    only  $(\mp 1, 0,\pm 1)$-type states are formed. For large strengths of SO coupling, supersolid-like superlattice and superstripe states are formed in the ferromagnetic phase.  In the  antiferromagnetic phase, for large strengths of SO coupling, 
supersolid-like superstripe and multi-ring states are formed. 
  For an equal mixture of     Rashba and Dresselhaus SO couplings,  only  a  superstripe state is found. 
 All these states are found to be dynamically stable and hence accessible in an    experiment and will 
enhance the  fundamental understanding of crystallization onto spatially periodic  states in solids.

\end{abstract}

 \maketitle
 
\section{1. Introduction}  

After the experimental realization \cite{bose} of a trapped Bose-Einstein condensate (BEC) of ultra-dilute alkali metal atoms at ultra-low temperature,  it has been possible to observe a hyperfine spin-1 ($F=1$) spinor BEC of $^{23}$Na atoms composed of the three spin components $F_z=\pm1, 0$ \cite{exptspinor}.  
The spin-orbit (SO) coupling naturally appears in  charged electrons and controls many properties of a solid including its crystalline or amorphous structure. 
Although, in a neutral atom, there cannot be a natural SO coupling,  it is  possible to introduce an artificial synthetic  SO coupling by  
Raman lasers that coherently couple the spin-component
states
in a spinor BEC 
\cite{thso}. 
Two  possible SO couplings are due to  Dresselhaus \cite{SOdre} and Rashba \cite{SOras}. 
An equal mixture of these   SO couplings  has been  realized  
in  a pseudo spin-1/2   $^{87}$Rb  \cite{exptso} and in a $^{23}$Na \cite{na-solid}  BEC  containing only two spin components   $F_z=0,-1$ of total spin $F=1$ and  in  
a spin-1 ferromagnetic  $^{87}$Rb BEC, containing all three spin components $F_z= \pm 1, 0$ \cite{exptsp1}. There has also been study of SO coupling in a fermionic system \cite{sherman}.
A spin-1  spinor BEC \cite{exptspinor}  appears in two  magnetic phases \cite{thspinorb} with distinct properties: ferromagnetic $(a_0>a_2)$ and  antiferromagnetic  $(a_0<a_2),$ where $a_0$ and $a_2$ are the scattering lengths in the total spin 0 and 2 channels, respectively.  
  
  A supersolid \cite{sprsld} is a  quantum state where
matter forms a  spatially-ordered stable structure as in a crystalline solid, breaking continuous transnational invariance,  and enjoys  friction-less flow as in a superfluid, breaking continuous gauge invariance. Hence a supersolid   simultaneously possesses qualities of a solid and a
superfluid, contrary to the intuition that friction-less
flow is a property exclusive to quantum fluids, e.g. Fermi superfluids \cite{fermi}, and 
BECs \cite{bose}. 
There have been theoretical suggestion for  supersolid-like states in an SO-coupled BEC \cite{solid-1/2}.
Recently, there has been experimental
confirmation of a periodic one-dimensional (1D) superstripe state with density modulation in an SO-coupled
pseudo spin-1/2 spinor BEC of  $^{23}$Na atoms \cite{st2} employing an equal mixture of Rashba and Dresselhaus couplings.
Following theoretical suggestions to create a supersolid  with dipolar \cite{losh} and finite-range \cite{finite} atomic interactions, more recently, different experimental groups also confirmed its presence in a dipolar   BEC in quasi-two-dimensional (quasi-2D)  \cite{dipolar2d} and quasi-1D  \cite{dipolar1d} geometries. 

In this paper, we demonstrate the appearance of  supersolid-like spatially-periodic states in  a harmonically-trapped quasi-2D 
Rashba or Dresselhaus SO-coupled spin-1  BEC in both ferromagnetic and  antiferromagnetic phases. In the ferromagnetic phase, the supersolid-like   state is an  excited state.
The ground state for all strengths $\gamma$ of SO coupling  in the ferromagnetic phase is  a circularly-asymmetric state without internal vortex. In addition, for  a small strength $\gamma$ $(\lessapprox 0.75)$ of SO coupling, one has the formation of  circularly-symmetric $(\mp 1,0,\pm 1)$-  and 
$(0,\pm 1,\pm 2)$-type states \cite{kita}, where the numbers in the parentheses denote the angular momentum of vortices at the center of the respective components $F_z=+1,0,-1$: the upper (lower) sign corresponds to Rashba (Dresselhaus) coupling and the negative sign denotes an anti-vortex of opposite vorticity.   For larger SO coupling ($\gamma \gtrapprox 0.75$), one encounters 
two types of supersolid-like  states in the ferromagnetic phase: (i) a state with an 1D periodic stripe in density and (ii) a
spontaneous crystallization onto  a 2D square-lattice state. In the 
 antiferromagnetic phase, there is no circularly-asymmetric state.
For a small $\gamma$  $(\lessapprox 0.75)$, one encounters a circularly-symmetric
$(\mp 1,0,\pm 1)$-type state \cite{kita} in the antiferromagnetic phase.  For larger SO coupling ($\gamma \gtrapprox 0.75$), two types of supersolid-like states are found: (i) a state with an 1D periodic stripe in density and (ii) a state with a radially-periodic    multi-ring pattern in density. For an equal mixture of Rashba and Dresselhaus  SO couplings, the only possible spatially-periodic state 
is the 1D stripe state  appearing in both ferromagnetic and   antiferromagnetic phases, in addition to a circularly-symmetric Gaussian-type state in the ferromagnetic phase; no other 
supersolid-like state is found. In the  ferromagnetic phase, the circularly-symmetric  state is the ground state. { These spatially-periodic states in an SO-coupled BEC,   sharing some properties of  a supersolid, 
  have their own names \cite{2020}, e.g., superstripe \cite{stringari}  and superlattice \cite{adhikari} states. 
A superstripe state has an 1D periodic pattern in density, whereas a superlattice state has a 2D periodic square-lattice pattern in density. }
In a quasi-1D and a quasi-2D uniform SO-coupled spin-1 spinor BEC, superstripe \cite{ad2}  and superlattice  \cite{adhikari} states have also been predicted.

In section 2 we present the mean-field  model  which we solve numerically to study the 
formation of spatially-periodic states in a harmonically-trapped  SO-coupled spin-1 BEC. In section 3
we present  numerical results for the formation of spatially-periodic states in a quasi-2D 
SO-coupled spin-1 BEC for Rashba, Dresselhaus couplings, and for an equal mixture of these couplings, employing  imaginary-time propagation. 
The dynamical stability of the spatially-periodic states are established by real-time propagation using the converged imaginary-time wave function as the initial state. Finally, in section 4 a summary of the findings is presented.

\section{2. Mean-field model for SO-coupled spin-1 BEC}
  
\label{II}

We consider a BEC of $N$ atoms, each of mass $ M$, under a quasi-2D
harmonic trap $V({\bf r})=    M\omega^2(x^2+y^2)/2+ M\omega_z^2 z^2/2$  $(\omega_z \gg \omega )$   , where   $\omega$ and $\omega_z$ are the angular trap frequencies in the $x-y$ plane and in the $z$ direction, respectively. The single-particle Hamiltonian of the three-dimensional (3D)  SO-coupled   BEC  is \cite{thspinorb}
%
\begin{equation}
H_0 = -\frac{\hbar^2}{2 M}(\nabla_{\boldsymbol \rho}^2+\partial_z^2)+\frac{ M}{2}[ \omega^2(x^2+y^2) +\omega_z^2 z^2]+ H_{\mathrm{SO}},
\label{sph} 
\end{equation}
where ${\boldsymbol \rho}\equiv \{x,y   \}$, $\nabla_{\boldsymbol \rho}^2\equiv 
(\partial_ x^2+\partial_ y^2)$,   $\partial_x \equiv \partial/\partial x$,   $\partial_y \equiv \partial/\partial y$,  
$\partial_z \equiv \partial/\partial z$. 
The SO-coupling interaction is \cite{exptso,zhai}
\begin{equation}
H_{\mathrm{SO}}= \gamma (\nu  p_y \Sigma_x-p_x \Sigma_y),
\end{equation}
 where 
$\nu=+1,-1,0$, respectively,  for Rashba and  Dresselhaus SO couplings, and  an  equal mixture of these SO couplings, 
 $\gamma$ is the strength of SO coupling,  
the $x$ and $y$ components of the momentum operator, respectively, 
 $p_x =-i\hbar \partial_x, p_y=-i\hbar \partial_y,$ and the irreducible representations of the $x$ and $y$ components of the  spin matrices  $\Sigma_x$ and $\Sigma_y$ are
\begin{eqnarray}
\Sigma_x=\frac{1}{\sqrt 2} \begin{pmatrix}
0 & 1 & 0 \\
1 & 0  & 1\\
0 & 1 & 0
\end{pmatrix}, \quad  \Sigma_y=\frac{i}{\sqrt 2 } \begin{pmatrix}
0 & -1 & 0 \\
1 & 0  & -1\\
0 & 1 & 0
\end{pmatrix}.
\end{eqnarray}

The reduced quasi-2D  coupled Gross-Pitaevskii (GP) equation \cite{sala} of 
the three spin components,  
for the 
SO-coupled spin-1 spinor  BEC, is \cite{thspinorb}
\begin{align}\label{EQ1} 
i \partial_t &\psi_{\pm 1}({\boldsymbol 
\rho})= \left[{\cal H}+{c_2}
\left(n_{\pm 1} -n_{\mp 1} +n_0\right)  \right] \psi_{\pm 1}({\boldsymbol 
\rho})\nonumber \\
+&\left\{c_2 \psi_0^2({\boldsymbol \rho})\psi_{\mp 1}^*({\boldsymbol \rho})\right\} 
-i {\widetilde \gamma} (\nu \partial_y \pm i \partial_x)  \psi_{0}  ({\boldsymbol \rho})  \, , 
\\ \label{EQ2}
i \partial_t&\psi_0({\boldsymbol \rho})=\left[ {\cal H}+{c_2}
\left(n_{+ 1}+n_{- 1}\right) \right] \psi_{0}({\boldsymbol \rho}) + \big \{2c_2 \psi_{+1}({\boldsymbol \rho})\nonumber \\
\times &\psi_{-1}({\boldsymbol \rho})\psi_{0}^* ({\boldsymbol \rho})\big\}   
-i{\widetilde \gamma} [-i  \partial_x   \phi_{-}  ({\boldsymbol \rho})  
 + \nu  \partial_y \phi_{+} ({\boldsymbol \rho}) ]   \, , \\
{\cal H}=&-\frac{1}{2}\nabla^2_{\boldsymbol \rho}+ \frac{1}{2}\rho^2  + c_0 n,    \\
c_0 =&\frac{2N\sqrt{2\pi \kappa}(a_0+2a_2)}{3},  \, \,  c_2 
= \frac{2N\sqrt{2\pi \kappa}(a_2-a_0)}{3}, \label{EQ4}
\end{align}
where $\kappa=\omega_z/\omega $, $\phi_{\pm} ({\boldsymbol \rho}) =\psi_{+1} ({\boldsymbol \rho}) \pm  \psi_{-1}({\boldsymbol \rho})$,
   $\partial_t \equiv \partial/\partial t$,   
 $\widetilde \gamma = \gamma/\sqrt 2 $, $n_j = |\psi_j|^2, j=\pm 1,0$ are the densities of spin components $F_z= \pm 1 , 0$, $n ({\boldsymbol \rho})= \sum_j n_j({\boldsymbol \rho})$  the total density,  
  and the asterisk denotes complex conjugate. All quantities in (\ref{EQ1})-(\ref{EQ4}) and in the following are dimensionless; this is achieved by  expressing lengths ($a_0,a_2,x,y,z$)
 in units of   oscillator length  
$l_0\equiv \sqrt{\hbar/ M\omega}$,
 energy in units of $\hbar \omega$, density  in units of $l_0^{-2}$, and time in units of $\omega^{-1}$.
In  ferromagnetic and   antiferromagnetic phases $c_2<0$, and $c_2>0$, respectively, and the normalization 
condition is ${\textstyle \int}n({\boldsymbol \rho})  \, d^2{ \rho}=1$.
The single-particle quasi-2D Hamiltonian corresponding to (\ref{EQ1}) and (\ref{EQ2})  is given by
\begin{equation}
\label{spar}
H_{\mathrm{2D}}= -\frac{1}{2}\nabla^2_{\boldsymbol \rho}+ \frac{1}{2}\rho^2+H_{\mathrm{SO}}. 
\end{equation}


The energy functional   for the time-independent version of 
  (\ref{EQ1})-(\ref{EQ2}) is 
   \begin{align}\label{energy}
E[\psi] &=  \textstyle \frac{1}{2} {\textstyle \int} d^2{ \rho} \big[ {\textstyle \sum}_j |\nabla_{\boldsymbol \rho}\psi_j|^ 2  + \rho^2 n+ 
c_0n^2\nonumber \\  &+ c_2\big\{n_{+1}^2 +n_{-1}^2
+2(n_{+1}n_0+n_{-1}n_0-n_{+1}n_{-1}\nonumber \\ &  +\psi_{-1}^*\psi_0^2\psi_{+1}^*  
+ \psi_{-1}\psi_0^{*2}\psi_{+1})  \big\} -2i\widetilde \gamma\big\{ \nu\psi_0^* 
 \partial_y \phi_{+}\nonumber \\
&
+ \nu\phi_{+}^* \partial_y \psi_0 
-i \psi_0^* \partial_x \phi_{-}
+i \phi_{-}^*\partial_x \psi_0 \big\}  \big] .
\end{align} 

\section{3. Numerical Results} 

\label{III}
 
We solve  (\ref{EQ1}) and (\ref{EQ2}) numerically by time propagation, after a discretization 
using the split-time-step Crank-Nicolson discretization rule \cite{bec2009} employing a space
step of 0.1 and a time step of 0.001 for imaginary-time propagation and 0.00025 for real-time propagation. 
The  lowest-energy stationary solutions of each type of symmetry are obtained by imaginary-time propagation.
The dynamical stability of a state is then tested by  real-time
propagation employing the converged imaginary-time wave function as the initial state. 
The normalization is held fixed during time propagation.
Magnetization $m=\int d^2 \rho [n_{+1}({\boldsymbol \rho})-n_{-1}({\boldsymbol  \rho})]$ is not  a conserved quantity and is allowed
to evolve freely during time propagation to attain a final converged value.

Instead of presenting the results in dimensionless units, we prefer to relate the results with the 
most commonly used  
ferromagnetic 
$^{87}$Rb and  antiferromagnetic $^{23}$Na atoms  with  the following scattering lengths, respectively: for  $^{87}$Rb    \cite{kokk} $a_0= 101.8a_B   $, $a_2 = 100.4a_B  $ and for $^{23}$Na \cite{crube}  $a_0= 50.0a_B$,
and $a_2 = 55.0a_B $, with $a_B$ $(=5.291772\times 10^{-11}$ m) the Bohr radius. In the ferromagnetic 
phase we present results for a harmonically trapped quasi-2D  spin-1 $^{87}$Rb  BEC of $N=10^5$ atoms with $ l_0/\sqrt{\kappa} = 2$ $\mu$m 
so that $c_0 \equiv 2N \sqrt{2\pi\kappa}(a_0+2a_2)/3l_0  \approx 1338$ and $c_2 \equiv  2N\sqrt{2\pi\kappa}(a_2-a_0)/3l_0  \approx -6.2$.   In the  antiferromagnetic 
phase we consider  a harmonically trapped quasi-2D  spin-1 $^{23}$Na  BEC of $N=1.891 \times 10^5$ atoms  
so that $c_0 \equiv 2N \sqrt{2\pi\kappa}(a_0+2a_2)/3l_0  \approx 1338$ and $c_2 \equiv  2N\sqrt{2\pi\kappa}(a_2-a_0)/3l_0  \approx 41.8$.   The number of atoms are so chosen so as to have the same $c_0$ in both cases.
In this fashion we keep
 $c_0$ fixed at 1338 
and change $c_2$ continuously from $-6.2$ to 41.8 to cover both ferromagnetic and  antiferromagnetic phases from  $^{87}$Rb to  $^{23}$Na.
{ This choice will allow a unified description of the ferromagnetic   $^{87}$Rb
and antiferromagnetic $^{23}$Na phases, viz. Figs. \ref{fig1}, \ref{fig2}(m)-(o),\ref{Fig4}(g)-(i), etc.
}
  
 {

\begin{figure}[!t] 
\centering
\includegraphics[width=\linewidth]{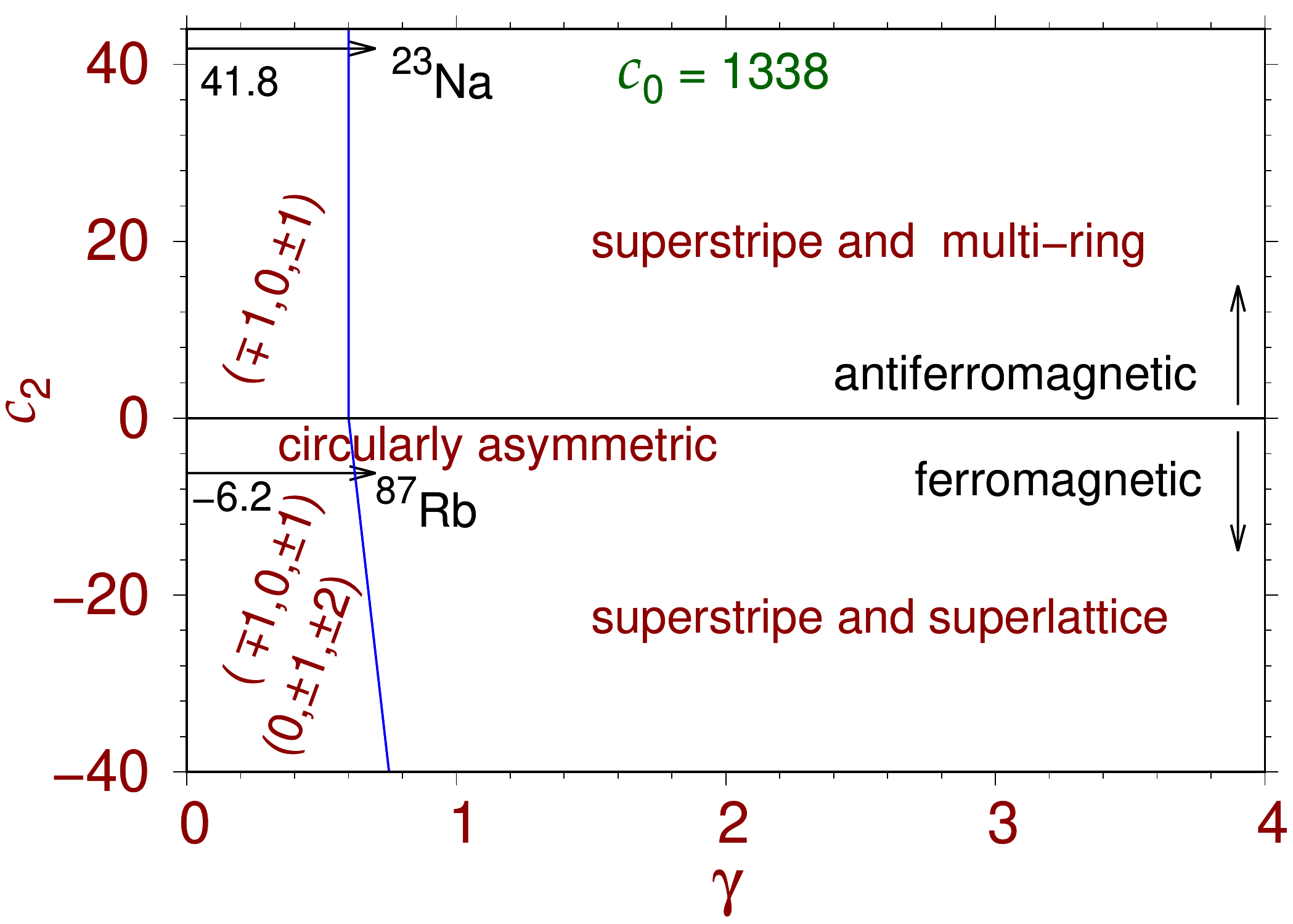}  
 
\caption{ The $c_2$ versus $\gamma$  phase plot for Rashba and Dresselhaus SO couplings
showing different types of possible  states in different regions of parameter space for $c_0=1338$.   Results in all figures are in dimensionless units. }
\label{fig1}

\end{figure}
}

\begin{figure}[!t] 
\centering
\includegraphics[width=.325\linewidth]{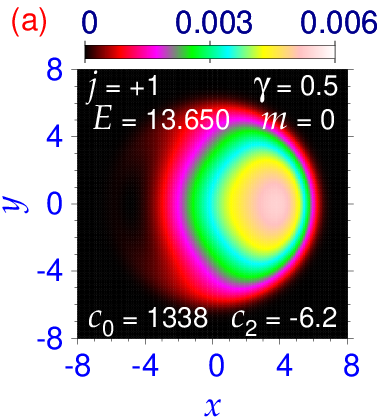} 
\includegraphics[width=.325\linewidth]{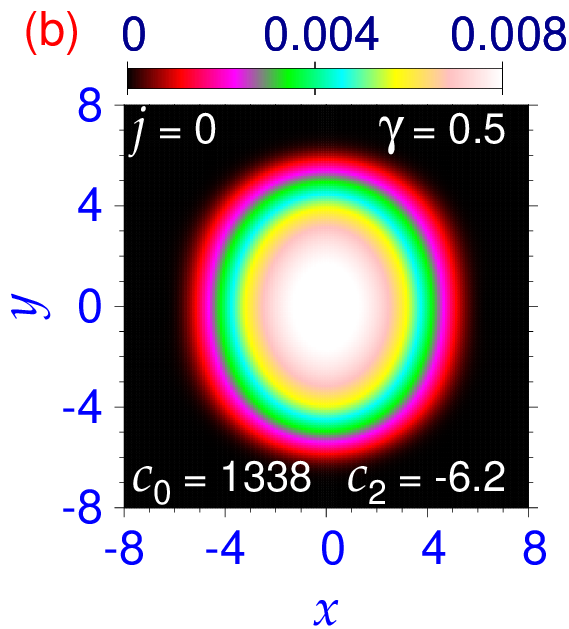}
\includegraphics[width=.325\linewidth]{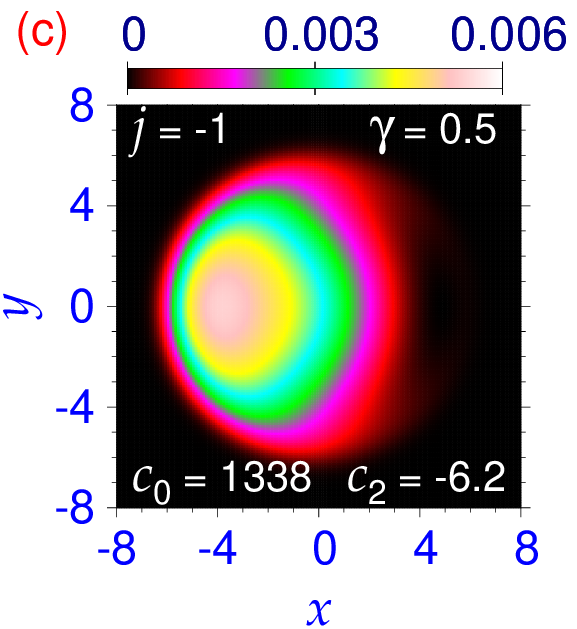}
 \includegraphics[width=.325\linewidth]{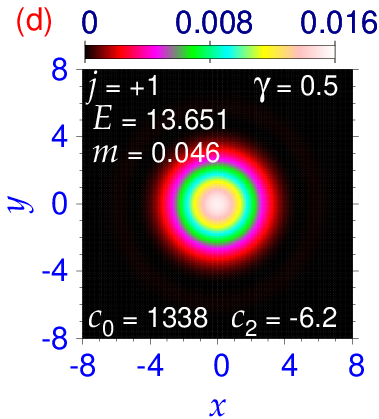} 
\includegraphics[width=.325\linewidth]{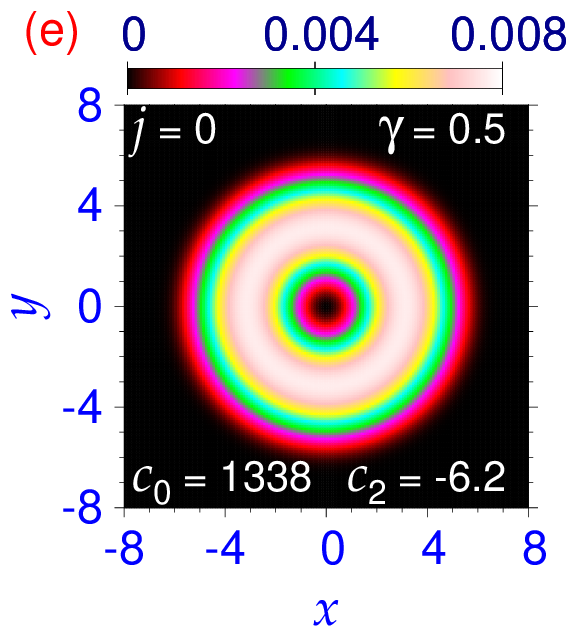}
 \includegraphics[width=.325\linewidth]{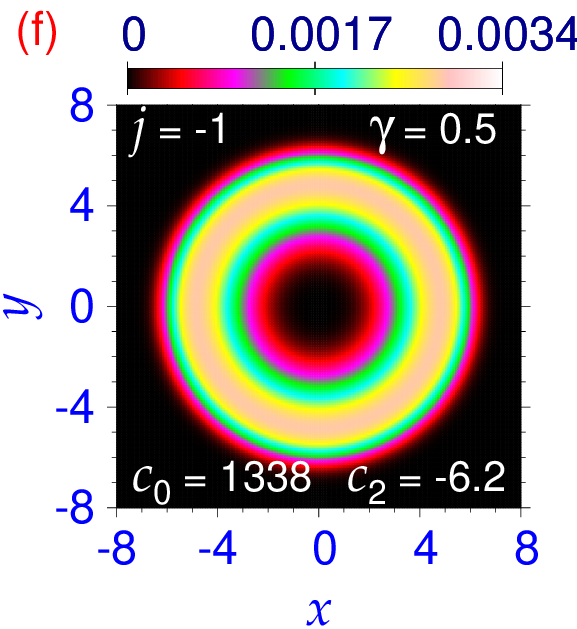}
 \includegraphics[width=.325\linewidth]{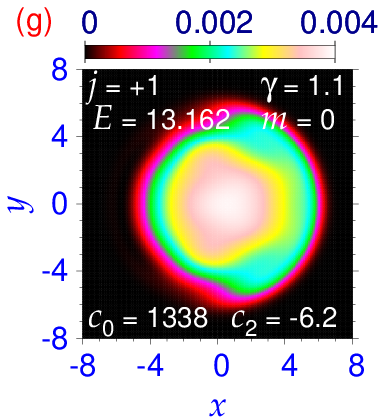} 
\includegraphics[width=.325\linewidth]{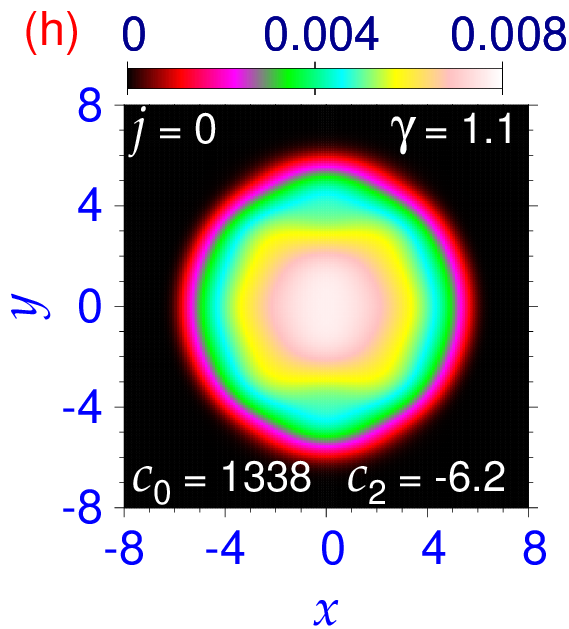}
\includegraphics[width=.325\linewidth]{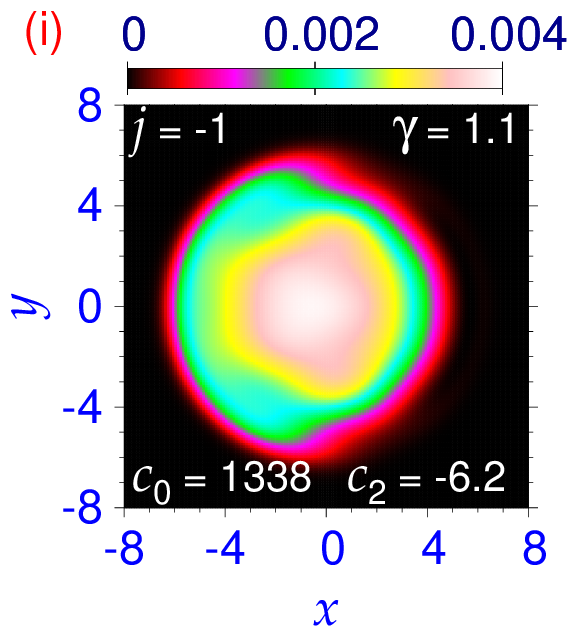}
 \includegraphics[width=.325\linewidth]{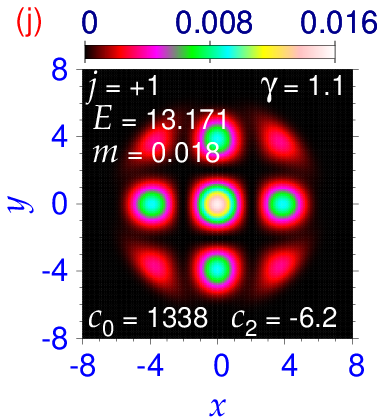} 
\includegraphics[width=.325\linewidth]{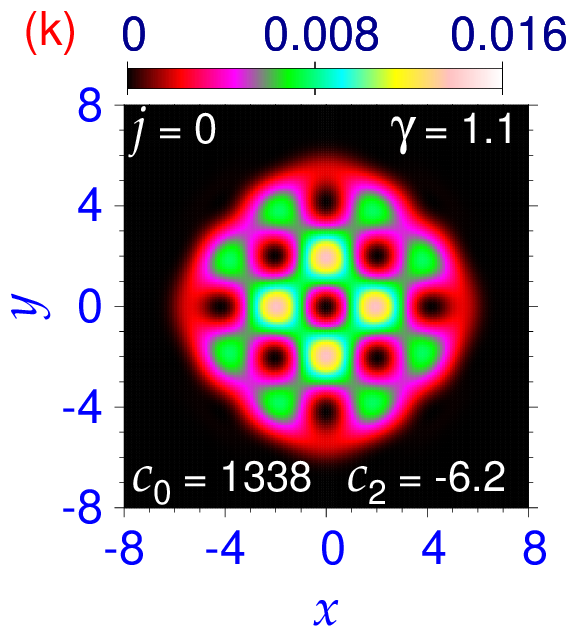}
\includegraphics[width=.325\linewidth]{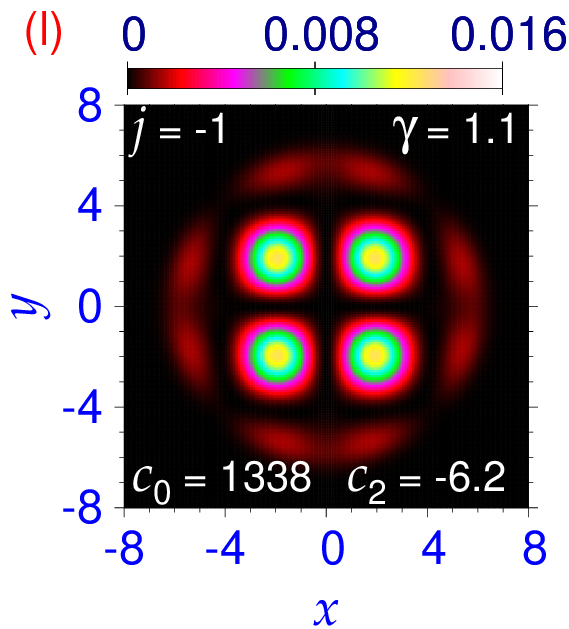}
 \includegraphics[width=.325\linewidth]{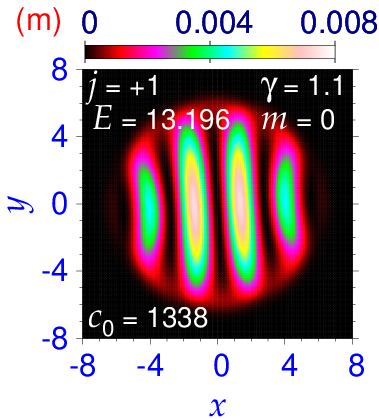} 
\includegraphics[width=.325\linewidth]{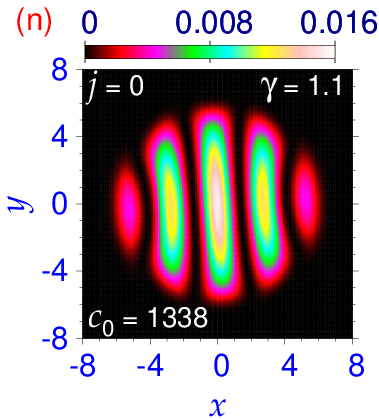}
\includegraphics[width=.325\linewidth]{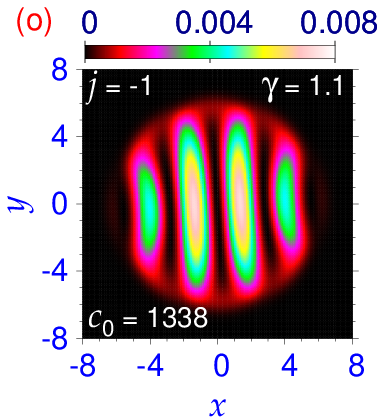}

\caption{Contour plot of density  $n_j(\boldsymbol  \rho)$ of a circularly-asymmetric 
Rashba or Dresselhaus SO-coupled ferromagnetic  $^{87}$Rb BEC for $\gamma=0.5$
of components  (a) $j=+1$, (b) $j=0$ and  (c)  $j=-1$ ; the same of a circularly-symmetric 
$(0,\pm 1,\pm 2)$-type Rashba or Dresselhaus SO-coupled
BEC for $\gamma=0.5$ in  (d) $j=+1$, (e) $j=0$ and  (f)  $j=-1$ ; 
the same of a circularly-asymmetric 
Rashba or Dresselhaus SO-coupled ferromagnetic  $^{87}$Rb BEC
for $\gamma=1.1$ 
of components  (g) $j=+1$, (h) $j=0$ and  (i)  $j=-1$ ;
the same  of a superlattice  SO-coupled ferromagnetic  $^{87}$Rb 
BEC for $\gamma =1.1,$
in   (j)-(l); the same  of a superstripe  SO-coupled ferromagnetic  $^{87}$Rb 
or antiferromagnetic  $^{23}$Na
BEC for $\gamma =1.1,$
in   (m)-(o).  The superstripe phase presented in 
(m)-(o) is independent of $c_2$.
The energy $E$ and magnetization $m$ are displayed in the plot of the $j=+1$ component in figures \ref{fig2}, \ref{Fig4},  \ref{fig4}, and \ref{fig5}. }
\label{fig2}

\end{figure}

The scenario of the formation of different types of states $-$ a circularly asymmetric state, circularly-symmetric  $(0,\pm 1, \pm 2)$- and   $(\mp 1,0, \pm 1)$-type states,  superstripe, multi-ring, and superlattice states  $-$    
 for Rashba and Dresselhaus SO couplings in the ferromagnetic and antiferromagnetic phases
is illustrated in figure  \ref{fig1} through a $c_2-\gamma$ phase plot for $c_0 = 1338$.  
In the ferromagnetic phase ($c_2<0$),  for all values of SO-coupling strength $\gamma$, the lowest-energy state is a circularly-asymmetric localized state  without any internal vortex. 
In addition, for a small $\gamma$ ($\gamma \lessapprox 0.75$) one can have either 
a circularly-symmetric $(0,\pm 1, \pm 2)$-type  state or 
 an excited   circularly-symmetric  $(\mp 1,0, \pm 1)$-type state.
For large $\gamma$, in the ferromagnetic phase  we have  either a superlattice state with 
periodic square-lattice pattern in density or a superstripe state with an 1D stripe in density. In the  antiferromagnetic phase ($c_2>0$), there is no circularly-asymmetric localized state, and 
for a small $\gamma$ ($\gamma \lessapprox 0.75$), one can have a
circularly-symmetric $(\mp 1,0, \pm 1)$-type  state. 
 For large $\gamma$, in the antiferromagnetic phase,  in addition to the 
1D superstripe state,     it is possible to have a  multi-ring 
state with radially-periodic   multi-ring pattern in density.

 We study the formation of the different types of states by a numerical solution of  (\ref{EQ1}) and
(\ref{EQ2}), by imaginary-time propagation,  in a Rashba {or Dresselhaus SO-coupled ferromagnetic 
$^{87}$Rb BEC with $c_0=1338$ and $c_2=-6.2$. For  $\gamma=0.5$, we find the quasi-degenerate circularly-asymmetric state as the ground state and the circularly-symmetric $(0,\pm 1,\pm 2)$-type state.  In figures \ref{fig2}(a)-(c) we illustrate the circularly-asymmetric 
state   
through a contour plot of density of the components $j=+1,0$ and  $j=-1$ for $\gamma=0.5$, respectively, obtained by using a localized state without vortices as the initial wave function.   The energy $E$ and magnetization $m$ of the states are displayed in the  contour density plot of component $j=+1$.  The numerical value of magnetization of all states presented in this paper are zero except the states in figures \ref{fig2}(d)-(f)
and \ref{fig2}(j)-(l).
In figures \ref{fig2}(d)-(f), we display  the component densities of the circularly-symmetric 
  $(0,\pm 1,\pm 2)$-type state for $\gamma=0.5$. To simulate the  $(0,\pm 1,\pm 2)$-type state,
in imaginary-time propagation, the vortices of angular momenta {$\pm 1$ and $\pm 2$} were imprinted in components $j=0$ and  $j=-1$  of the localized initial wave function. The central hole in components $j=0, -1$ of figures \ref{fig2}(e)-(f) corresponds to the {core of vortices/anti-vortices of angular momenta $\pm 1, \pm 2$},
respectively.
Although the densities of the Rashba and Dresselhaus SO-coupled states are the same, their phases are different. The contour plot of the phase of the wave function components $j=+1,0,-1$ of the Rashba SO-coupled state of figures \ref{fig2}(a)-(c)
 is shown in figures \ref{fig3}(a)-(c); the same of the Dresselhaus SO-coupled state  is shown in 
figures \ref{fig3}(d)-(f). The phase drops upon  a complete rotation in plots \ref{fig3}(b)-(c) for Rashba SO coupling 
are  $2\pi$ and 
$4\pi$ indicating angular momenta of +1 and +2 in these components. The corresponding angular momenta 
of  components $j=0$ and  $j=-1$  from figures \ref{fig3}(e)-(f)
for Dresselhaus SO coupling are $-1$ and $-2$.   
The energies of the circularly-asymmetric and the $(0,\pm 1,\pm 2)$-type states for $\gamma=0.5$ shown in  figures \ref{fig2}(a)-(c) and \ref{fig2}(d)-(f)  
are $E=13.650$ and 13.651, respectively. In this case there is also a $(\mp 1,0,\pm 1)$-type excited state with energy $E=13.681$, appearing in both ferromagnetic and antiferromagnetic phases, viz. figures  \ref{fig4}(a)-(c).

\begin{figure}[!t] 
\centering
\includegraphics[width=.308\linewidth]{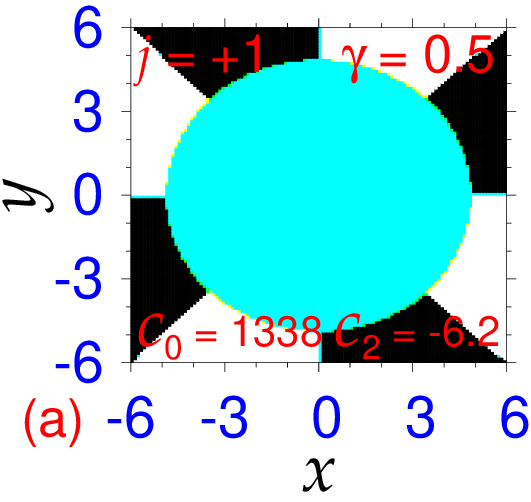} 
\includegraphics[width=.308\linewidth]{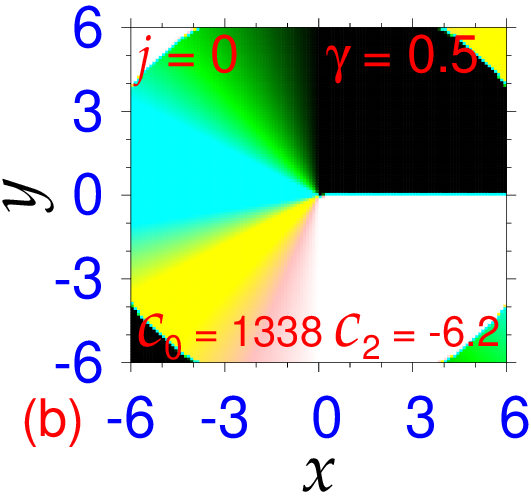}
\includegraphics[width=.359\linewidth]{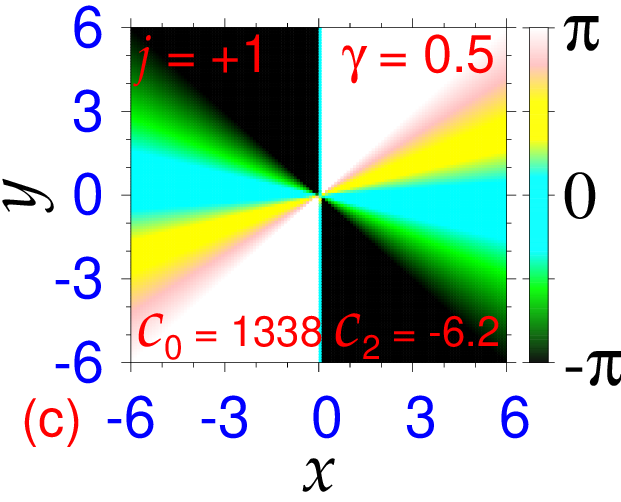}
 \includegraphics[width=.308\linewidth]{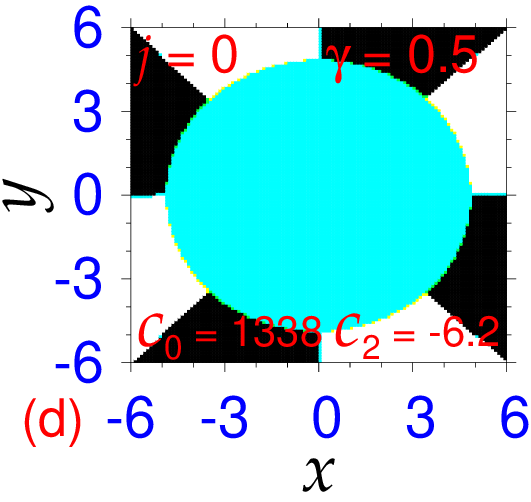} 
\includegraphics[width=.308\linewidth]{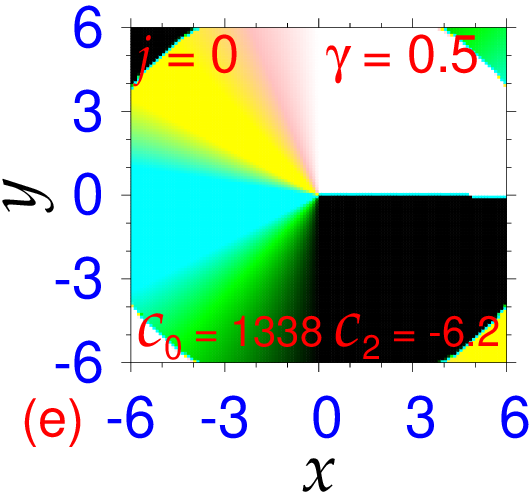}
\includegraphics[width=.359\linewidth]{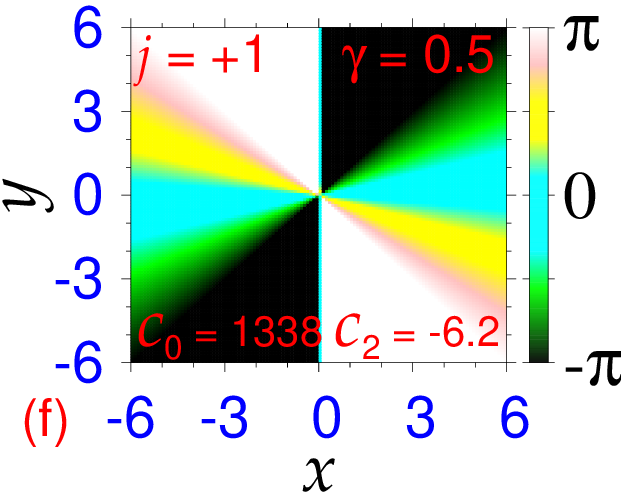}
 \includegraphics[width=.308\linewidth]{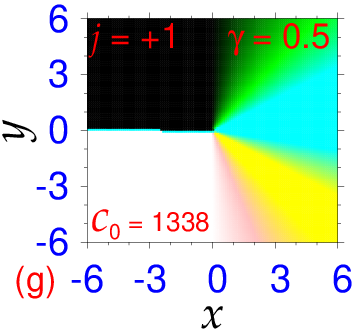} 
\includegraphics[width=.308\linewidth]{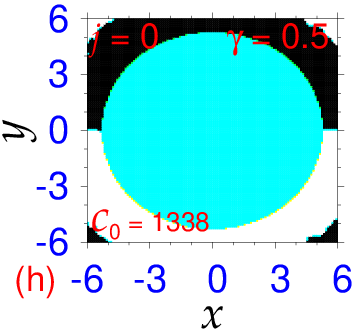}
\includegraphics[width=.359\linewidth]{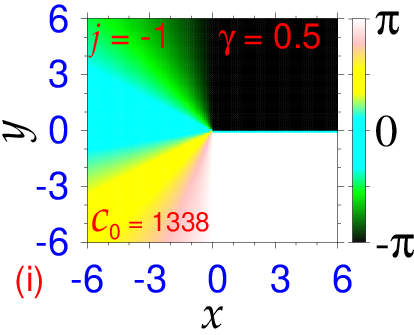}
 \includegraphics[width=.308\linewidth]{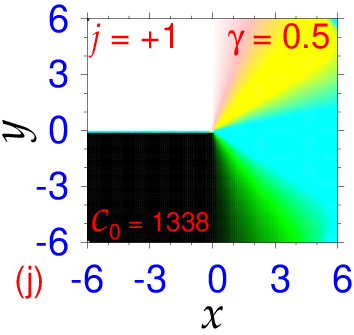} 
\includegraphics[width=.308\linewidth]{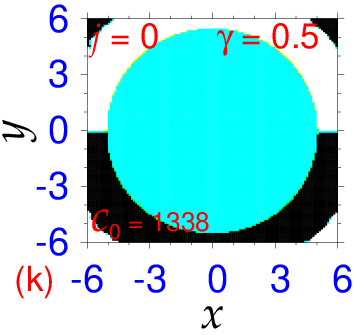}
\includegraphics[width=.359\linewidth]{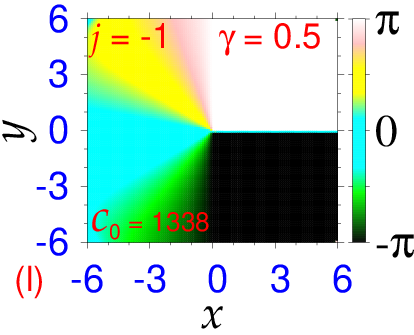} 
\

\caption{The contour plot of the phase of the wave function of the $(0,+1,+2)$-type 
Rashba SO-coupled ferromagnetic BEC of figures \ref{fig2}(d)-(f) of components  (a) $j=+1$, (b) $j=0$, and
(c)  $j=-1$ ;  the same of the $(0,-1,-2)$-type Dresselhaus  SO-coupled ferromagnetic BEC of figures 
\ref{fig2}(d)-(f) of components  (d) $j=+1$, (e) $j=0$, and
(f)  $j=-1$ ; the same of the $(-1,0,+1)$-type  
Rashba SO-coupled BEC of figures  \ref{fig4}(a)-(c)  of components  (g) $j=+1$, (h) $j=0$ and (i)  $j=-1$ ;
the same of a $(+1,0,-1)$-type  
Dresselhaus SO-coupled BEC figures  \ref{fig4}(a)-(c) of components  (g) $j=+1$, (h) $j=0$ and (i)  $j=-1$ . 
}
\label{fig3}

\end{figure}

As $\gamma$ is increased, the circularly-asymmetric state continues to exist in the ferromagnetic phase as the ground state, but the $(0,\pm 1,\pm 2)$-type  state
evolves into a superlattice state with a periodic square-lattice pattern in density;
the  $(\mp 1,0,\pm 1)$-type  state simply cease to exist.
{The single-particle Hamiltonian (\ref{spar}) should have solutions of  the plane wave form $\exp(\pm i \gamma x)\times \exp (\pm i \gamma y)$ or of the form $\exp (\pm i \gamma \rho)$. In  the presence of interaction ($c_0,c_2 \ne 0$), the solution will be a superposition of such plane wave solutions leading to a periodic variation of density in the form $\sin^2( \gamma x)$, $\cos^2( \gamma y)$, $\sin^2( \gamma x)\sin^2( \gamma y)$,
 $\sin^2( \gamma \rho)$ etc. appropriate for superstripe, superlattice or multi-ring BECs, respectively. This has been demonstrated in detail for quasi-1D SO-coupled BEC \cite{quasi-1d1}, the same of  the present quasi-2D SO-coupled BEC will be the subject of a future investigation. Although superstripe, superlattice and multi-ring states are possible from a consideration of the mean-field GP equation, for certain parameter domain one of these becomes the ground state, the others could appear as excited states or could even be unstable. For example, we will see that,  in the  antiferromagnetic phase, the superstripe state is the ground state, the multi-ring state is an excited state, viz. figures \ref{Fig4}(g)-(i) and \ref{fig4}, and the superlattice state is unstable.  However, we established in  \cite{adhikari} that in a {\it uniform} spin-1 SO-coupled BEC the superlattice state can appear in both ferromagnetic as well as  antiferromagnetic phases. In the ferromagnetic phase, the circularly asymmetric state is the ground state, viz. figure \ref{fig2}.}

\begin{figure}[!t] 
\centering
\includegraphics[width=.325\linewidth]{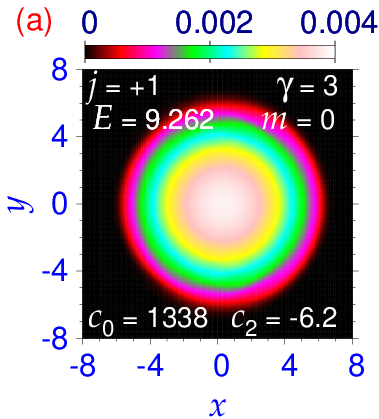} 
\includegraphics[width=.325\linewidth]{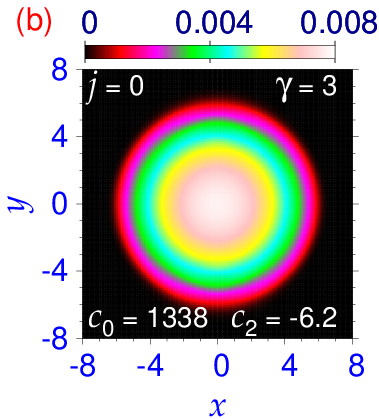}
\includegraphics[width=.325\linewidth]{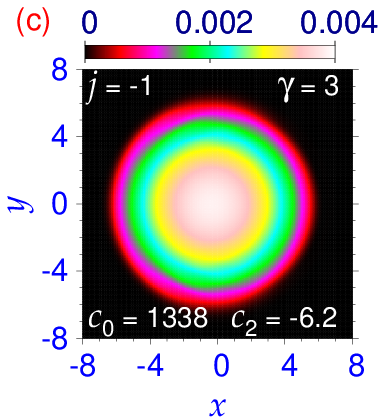}
 \includegraphics[width=.325\linewidth]{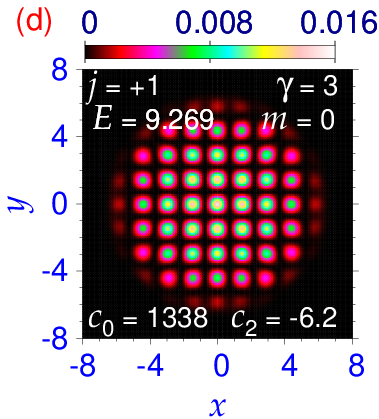} 
\includegraphics[width=.325\linewidth]{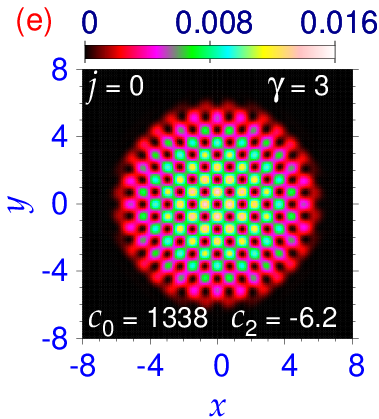}
\includegraphics[width=.325\linewidth]{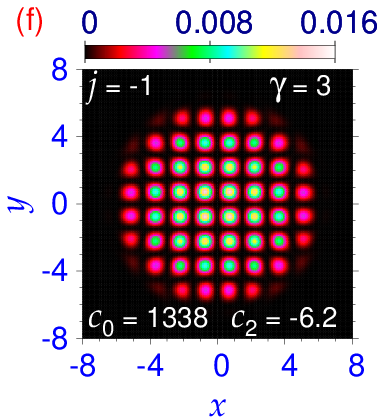}
 \includegraphics[width=.325\linewidth]{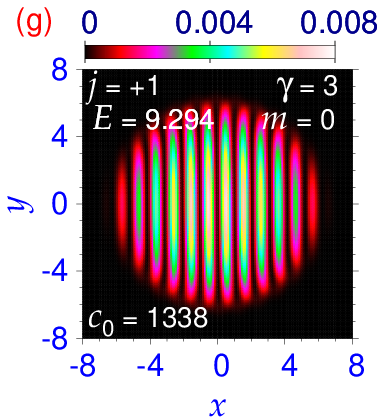} 
\includegraphics[width=.325\linewidth]{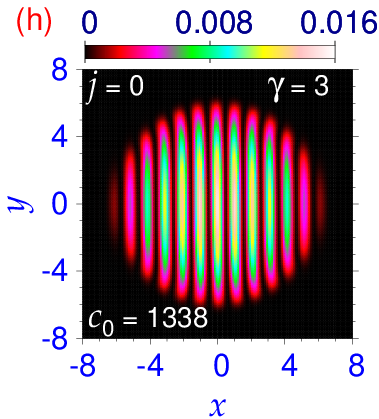}
\includegraphics[width=.325\linewidth]{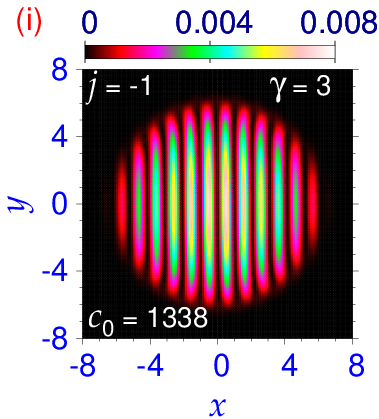} 

\caption{Contour plot of density  $n_j(\boldsymbol  \rho)$ of a circularly-asymmetric  
Rashba or Dresselhaus SO-coupled  ferromagnetic  $^{87}$Rb BEC for $\gamma=3$
of components  (a) $j=+1$, (b) $j=0$ and  (c)  $j=-1$ ; the same of a superlattice
Rashba or Dresselhaus SO-coupled
BEC for $\gamma=3$ of  components (d) $j=+1$, (e) $j=0$ and  (f)  $j=-1$ ; 
the same of a superstripe 
Rashba or Dresselhaus SO-coupled ferromagnetic  $^{87}$Rb  or an antiferromagnetic  
 $^{23}$Na 
BEC
for $\gamma=3$ 
of components  (g) $j=+1$, (h) $j=0$ and  (i)  $j=-1$ .  In  plots  (a)-(f) $c_0=1338, c_2=-6.2$;  the profile in plots (g)-(i) is for $c_0=1338$ and  independent of $c_2$ thus holding for both ferromagnetic and antiferromagnetic phases. }

\label{Fig4}

\end{figure}

Using the converged circularly-symmetric $(0,\pm 1,\pm 2)$-type 
wave function for $\gamma=0.5$ as the initial state in imaginary-time propagation, we find the superlattice  state for larger $\gamma$  ($\gamma \gtrapprox 0.75$), while  the circular symmetry 
is spontaneously broken and  the formation of superlattice  starts.  
The  circularly-asymmetric and superlattice  states 
for $\gamma=1.1$
are illustrated 
through a contour plot of densities of components $j=+1,0$ and  $j=-1$, respectively, in figures \ref{fig2}(g)-(i) and (j)-(l) with energies $E=13.162$ and 13.171. In addition, there is the excited superstripe state with energy  13.196 displayed in figures 
 \ref{fig2}(m)-(o). The energy and density of this superstripe state is independent of $c_2$ (in the range $40>c_2>-40$)
and hence is the same in both ferromagnetic and antiferromagnetic phases.  In the numerical simulation of the 
superstripe state by imaginary-time propagation, the stripe pattern is imprinted on the initial state as: $\psi_{\pm 1}({\boldsymbol \rho}) \sim \sin(\gamma x) \psi_{\mathrm{loc}} ({\boldsymbol \rho}),  \psi_0({\boldsymbol \rho}) \sim \cos(\gamma x) \psi_{\mathrm{loc}} ({\boldsymbol \rho})$, where $\psi_{\mathrm{loc}} ({\boldsymbol \rho})$ is a localized normalizable state.

With the increase of SO coupling $\gamma$ ($=3$), in the ferromagnetic phase the circularly asymmetric state continues as the ground state $(E=9.262)$, with reduced asymmetry, as shown in figures \ref{Fig4}(a)-(c), where we plot the  contour densities of components $j=+1,0,-1$, respectively. 
The superstripe state for the same parameters as displayed in figures \ref{Fig4}(d)-(f) for $\gamma =3$
$(E=9.269)$ is an excited state. In this state, the 
 square-lattice pattern has become more prominent with the increase of $\gamma$. 
For all $\gamma$, the square lattice for the $j=0$ component makes an
angle of 45$^{\circ}$ with the square lattice for $j=\pm 1$ components and  there is no 
symmetry between densities of $j=\pm 1$ components. 
 As $\gamma $ increases,  the array  of lattices in the spatially-periodic state is denser and the number of occupied sites 
  increases, although the size of the condensate  does not increase: the size is controlled by the 
external trap. 
The profile of the superstripe state with $\gamma=3,c_0=1338$ is shown in figures  \ref{Fig4}(g)-(i). The energy and density of this superstripe state  are independent of $c_2$; hence the results are valid for both ferromagnetic and antiferromagnetic phases.
However, the superstripe state $(E=9.294)$ is an excited state in the ferromagnetic phase, whereas it is the ground state in the antiferromagnetic phase.
 The result of densities and energies  for Rashba and Dresselhaus SO couplings are the same, although the underlying wave functions are different.

Now we consider the formation of a spatially-periodic state in an  antiferromagnetic  SO-coupled spin-1 $^{23}$Na BEC with 
$c_0=1338$ and $c_2=41.8$. For a small 
$\gamma$ ($\gamma \lessapprox 0.75$), the only possible state is a circularly-symmetric 
$(\mp 1,0,\pm 1)$-type state \cite{kita}, for Rashba and Dresselhaus SO couplings, respectively, as displayed in
the contour density plot of  figures \ref{fig4}(a)-(c)  for $\gamma=0.5$. In imaginary-time propagation, the vortices of angular momenta $\mp 1$ ($\pm 1$) were imprinted in components
$j =\pm 1 $ of the initial wave function. 
The energy ($E=13.681$)  and density of this state are independent of $c_2$;  hence the results are valid for both  antiferromagnetic and    ferromagnetic phases for weak SO coupling $(\gamma \lessapprox 0.75)$. 
However, the  $(\mp 1,0,\pm 1)$-type  state is an excited state in the ferromagnetic phase, whereas it is the ground state in the antiferromagnetic phase.
The phases of the wave function for  Rashba and Dresselhaus SO couplings are different. In figures \ref{fig3}(g)-(i) and \ref{fig3}(j)-(l) we illustrate the contour plot of the phases of the wave function components 
$j=+1,0,-1$ for the BECs of figures \ref{fig4}(a)-(c) for Rashba and Dresselhaus couplings, respectively. These phase plots confirm the angular momenta of $\mp 1$ and $ \pm 1$ in components $j=\pm 1$ for Rashba and Dresselhaus couplings, respectively.
 Using the converged wave function for $\gamma=0.5$ as the initial state, we obtain the multi-ring  state for larger values of $\gamma$, by increasing $\gamma$ slowly 
during imaginary-time propagation,
as illustrated in figures \ref{fig4}(d)-(f)  for $\gamma=3$ in the contour density plot of the components $j=+1,0,-1$. The anti-vortex and vortex at the center of the components 
$j=\pm 1$ of figures \ref{fig4}(a) and (c) survive in  figures \ref{fig4}(d) and (f). For
$\gamma=3$  the ground state   in the  antiferromagnetic phase is the superstripe state, which can be obtained by imaginary-time propagation using an  initial Gaussian state with  periodic stripe modulation, as shown in figures \ref{Fig4}(g)-(i). This state is an excited state in  the ferromagnetic phase.
In   $^{23}$Na with $c_0 =1338$ and $c_2=41.8$, 
of the two types of periodic states the superstripe   state ($E=9.294$),  viz. figure \ref{Fig4}(g)-(i), has the smaller energy than the multi-ring state ($E=9.344$), viz. figure \ref{fig4}(d)-(f). The energy of the multi-ring state is independent of the nonlinearity $c_2$ $(c_2>0)$.

 \begin{figure}[!t] 
\centering
\includegraphics[width=.325\linewidth]{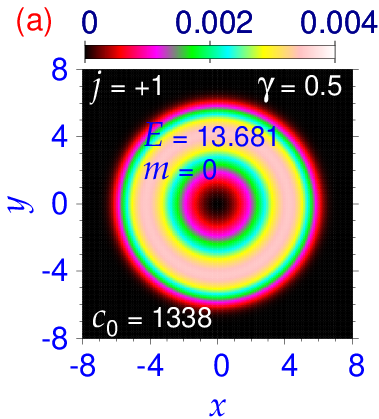} 
\includegraphics[width=.325\linewidth]{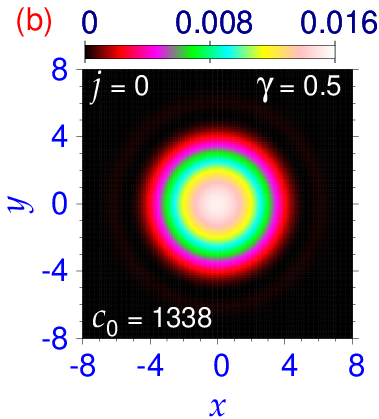}
\includegraphics[width=.325\linewidth]{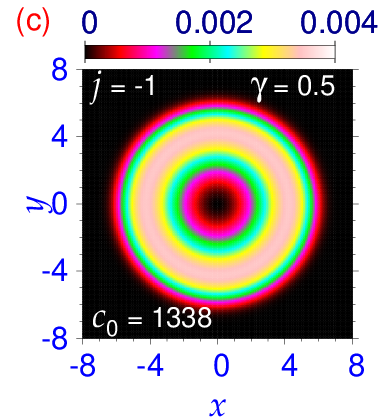}
 \includegraphics[width=.325\linewidth]{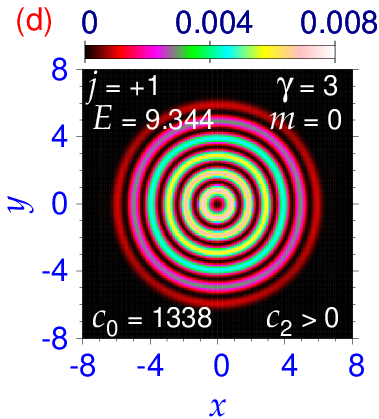} 
\includegraphics[width=.325\linewidth]{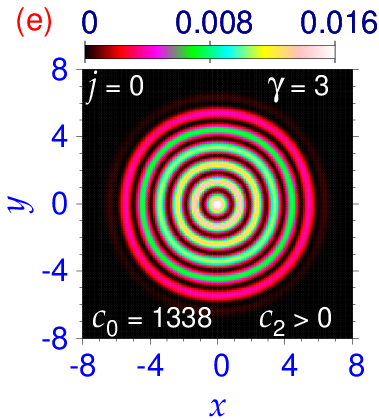}
\includegraphics[width=.325\linewidth]{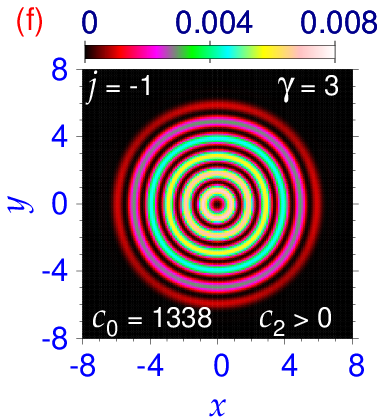}

\caption{Contour plot of density  $n_j(\boldsymbol  \rho)$ of a circularly-symmetric $(\mp 1,0, \pm 1)$-type 
Rashba or Dresselhaus SO-coupled  antiferromagnetic  $^{23}$Na 
or  ferromagnetic  $^{87}$Rb 
BEC for $\gamma=0.5$
of components  (a) $j=+1$, (b)$j=0$ and  (c)  $j=-1$ ; the same of a circularly-symmetric multi-ring
Rashba or Dresselhaus SO-coupled  antiferromagnetic  $^{23}$Na
BEC for $\gamma=3$ of  components (d) $j=+1$, (e)$j=0$ and  (f)  $j=-1$ .  In all plots $c_0=1338, c_2=41.8$}

\label{fig4}

\end{figure}

\begin{figure}[!t] 
\centering
\includegraphics[width=.325\linewidth]{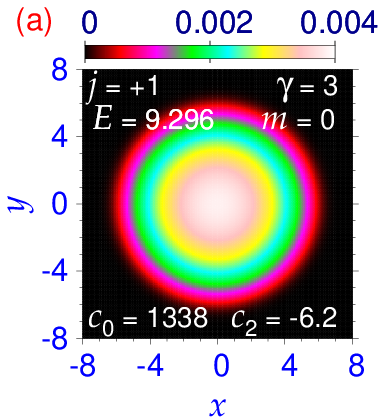} 
\includegraphics[width=.325\linewidth]{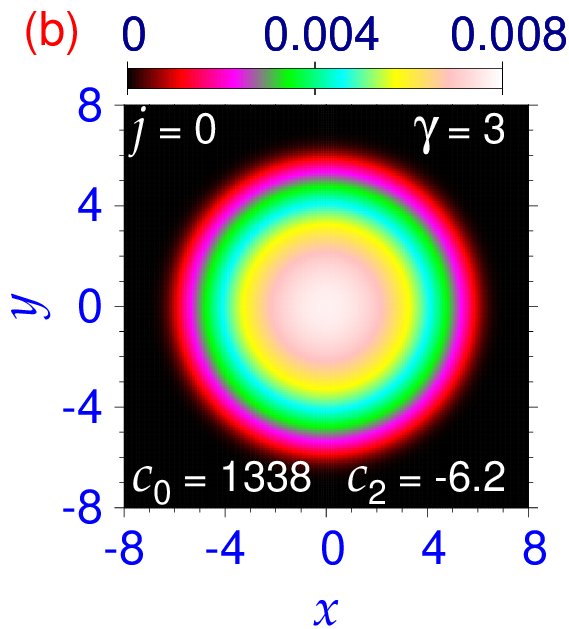}
\includegraphics[width=.325\linewidth]{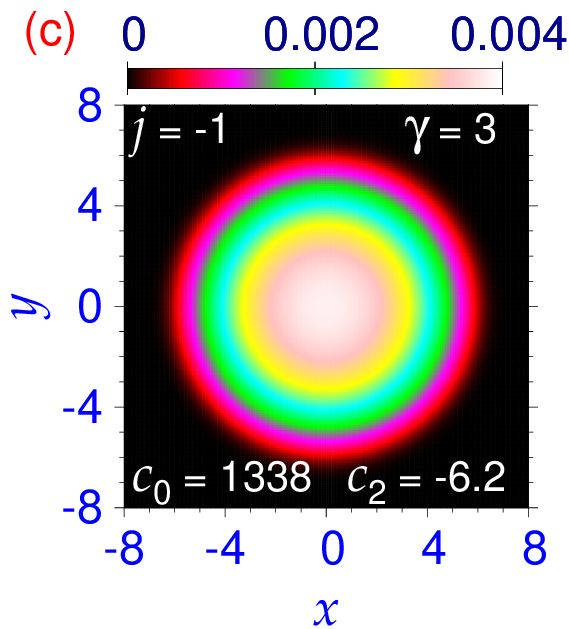}
 \includegraphics[width=.325\linewidth]{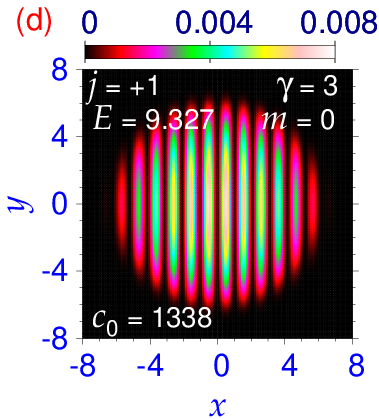} 
\includegraphics[width=.325\linewidth]{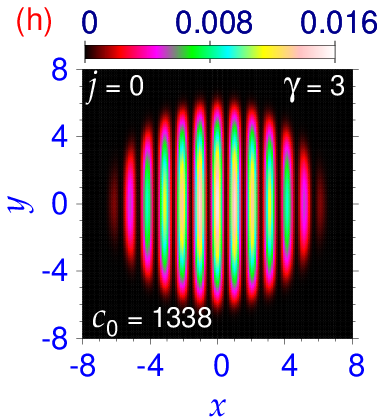}
\includegraphics[width=.325\linewidth]{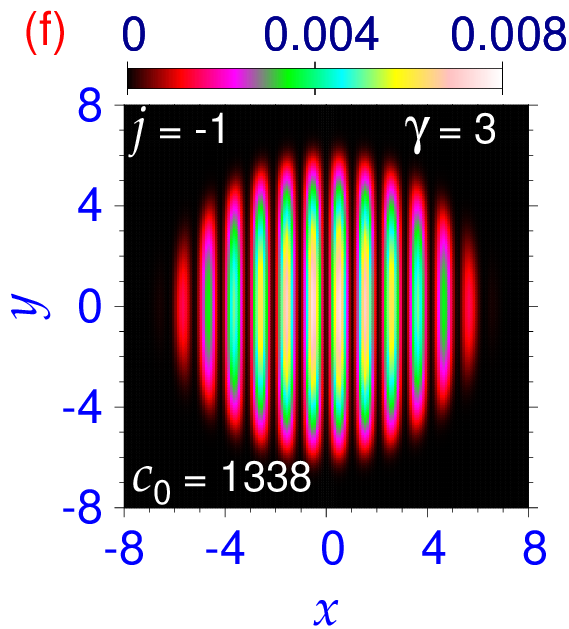}
 
\

\caption{ Contour plot of density  $n_j(\boldsymbol  \rho)$ of a circularly-symmetric
 ferromagnetic $^{87}$Rb BEC for an equal mixture of Rashba and Dresselhaus SO-couplings with $\gamma=3$
of components  (a) $j=+1$, (b) $j=0$ and  (c)  $j=-1$ ; the same of a superstripe ferromagnetic or  antiferromagnetic 
BEC for an equal mixture of Rashba and Dresselhaus SO-couplings with $\gamma=3$ of components 
   (d) $j=+1$, (e) $j=0$ and  (f)  $j=-1$ . }

\label{fig5}

\end{figure}

For an equal mixture of Rashba and Dresselhaus  SO couplings  a superstripe state is possible in both ferromagnetic and  antiferromagnetic phases \cite{solid-1/2,st2}. In the  antiferromagnetic phase the superstripe state is the only possible state. In the ferromagnetic phase, a circularly-symmetric vortex-free state appears as the ground state 
$(E=9.296)$
as shown through a contour plot of densities of components $j=+1,0,-1$ in figures \ref{fig5}(a)-(c) for $\gamma=3.$
The superstripe state is illustrated in    figures \ref{fig5}(d)-(f). The energy 
($E=9.327$)  and the density of the superstripe state is independent of 
$c_2$; it is an excited state in  the ferromagnetic phase.
\begin{figure}[!t] 
\centering
\includegraphics[width=.325\linewidth]{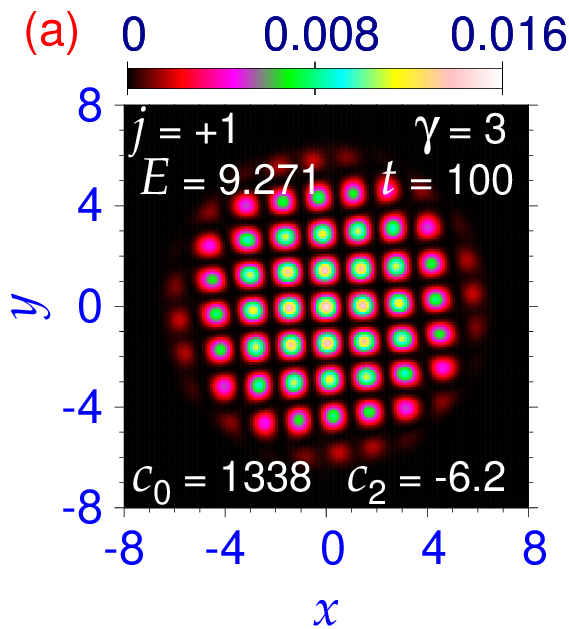} 
\includegraphics[width=.325\linewidth]{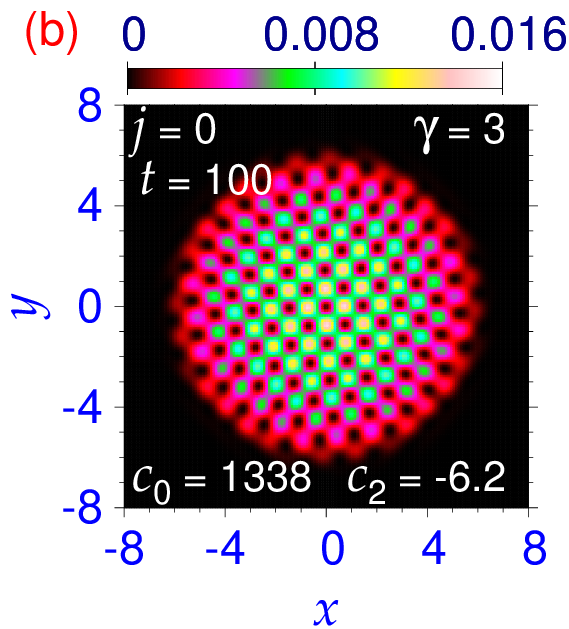}
\includegraphics[width=.325\linewidth]{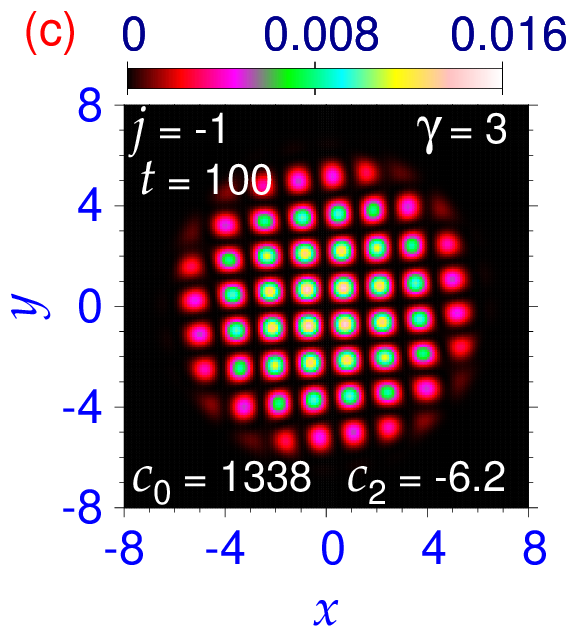}
 \includegraphics[width=.325\linewidth]{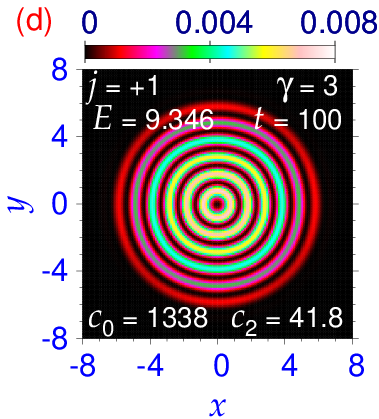} 
\includegraphics[width=.325\linewidth]{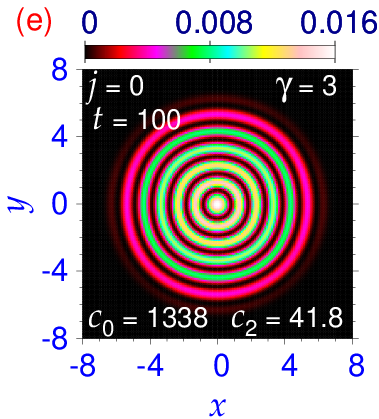}
\includegraphics[width=.325\linewidth]{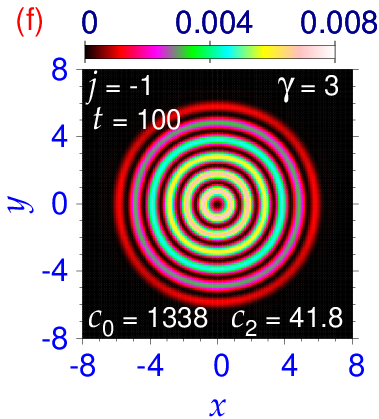}
\

\caption{ Contour plot of component densities of the SO-coupled BEC, of figures \ref{fig2}(m)-(o),
of  components (a) $j=+1$  (b) $j=0$   and (c)  $j=-1$ , after real-time propagation during 100 units of time;
the  same of the SO-coupled BEC,  of 
 figures \ref{fig4}(d)-(f),  of  components (d) $j=+1$  (e) $j=0$   and (f)  $j=-1$ , after real-time propagation during 100 units of time.  The initial state used in real-time propagation is the corresponding convergent wave function obtained in  imaginary-time propagation. }

\label{fig6}

\end{figure}

We test the dynamical stability of the SO-coupled states considered here via a real-time propagation over an extended period of time using the corresponding convergent wave function obtained in imaginary-time propagation as the initial function.  As an illustration we consider the superlattice state   of figures \ref{Fig4}(d)-(f) and  the  multi-ring  state of figures \ref{fig4}(d)-(f)
and subject the corresponding imaginary-time wave functions  to real-time propagation during 100 units of time.  The resultant densities for 
the components $j=+1,0,-1$  are displayed in figures \ref{fig6}(a)-(c), and (d)-(f)  respectively. 
Although the energy and size oscillate during time evolution, the periodic pattern in density survives at $t=100$, viz.
slightly different energies at $t=100$ ($E=9.271$ and 9.346) compared to the corresponding converged imaginary-time results ($E=9.269 $ and 9.344). 
 These densities are in good agreement with the corresponding imaginary-time  densities, demonstrating the dynamical stability of the spatially-periodic state.

\section{4. summary}

\label{IV}

We demonstrated new types of dynamically stable supersolid-like spatially-periodic states 
in a Rashba or a Dresselhaus SO-coupled   harmonically-trapped spin-1 quasi-2D   spinor  BEC  
for different SO-coupling strength $\gamma$, in both ferromagnetic and  antiferromagnetic phases,
using  a numerical solution of the GP equation.  In the ferromagnetic phase, the ground state for all strengths of SO coupling 
is a circularly-asymmetric localized state  without any internal vortex. 
 In the  antiferromagnetic phase,
there is no  circularly-asymmetric localized   state.
For small strengths of SO coupling ($\gamma \lessapprox 0.75$)
there is no spatially-periodic state.  For  small $\gamma$, for both Rashba and Dresselhaus SO couplings, in the ferromagnetic phase  $(0,\pm 1,\pm 2)$- and 
$(\mp 1,0, \pm 1)$-type 
states are found, whereas in the  antiferromagnetic phase {\it only}
a $(\mp 1,0, \pm 1)$-type state is found.   
In the  ferromagnetic phase, for large strengths of Rashba and Dresselhaus SO couplings $(\gamma \gtrapprox 0.75)$, the  possible spatially-periodic states   are  superlattice and superstripe  states, whereas,  in the  antiferromagnetic phase,  
 superstripe  and  multi-ring states are found. A superstripe state  was also suggested \cite{losh} and observed \cite{st2} in a 
pseudo spin-1/2 $^{23}$Na BEC. { There has been  detailed study \cite{mal} of different phases of a trapped SO-coupled quasi-2D pseudo spin-1/2 BEC.  The 
semi-vortex state explored there should in present notation  be a $(0, \pm 1)$-type state with a single vortex/antivortex in one component, quite similar to the  $(\mp 1, 0, \pm 1)$-type state of this paper. Although, no superlattice or superstripe state was reported in that paper \cite{mal},  a quasi-2D  SO-coupled trapped pseudo spin-1/2 BEC  is appropriate for hosting a superlattice or superstripe state. The search of such a  spatially-periodic state in this system is an interesting topic of future investigation.   A superlattice state was first predicted 
 in a quasi-2D SO-coupled  spin-1 uniform BEC \cite{adhikari}.}

For an equal mixture of Rashba and Dresselhaus couplings,  a superstripe state appears in both ferromagnetic and  antiferromagnetic phases in addition to a circularly-symmetric Gaussian type state in the ferromagnetic phase. In the ferromagnetic phase the ground state is the circularly-symmetric state.  
The dynamic stability of  these states was established by steady real-time propagation over a long period of time using the converged wave function of imaginary-time calculation as the initial state.  

It is remarkable that the superlattice states first found in a uniform system \cite{adhikari} survives  
under a strong harmonic trap.
These states  are dynamically robust and deserve  further  theoretical, and also experimental,  investigation. { The energy difference between different types of states, e.g. multi-ring state, superstripe state, superlattice state, etc., is very small  for experimental measurement.  Hence, for experimental purpose, these states for  a fixed set of parameters could be  considered degenerate.  Nevertheless, as all these states are dynamically stable,  in an experiment, depending on the initial conditions, any of these quasi-degenerate states  could be accessed. }
 A natural extension of this study is to search for spatially-periodic states  in an SO-coupled 3D spin-1 
and a quasi-2D spin-2
BEC, where a rich variety of spatially-periodic states are expected. { It would  also be of interest to study the role of the Zeeman-like terms  \cite{mal} on the present multi-ring, superlattice and superstripe  states.  }

\begin{acknowledgments}
S.K.A. acknowledges support by the CNPq (Brazil) grant 301324/2019-0, and by the ICTP-SAIFR-FAPESP (Brazil) grant 2016/01343-7

\end{acknowledgments}


\begin{thebibliography}{6}


\bibitem{bose}M. H. Anderson, J. R. Ensher, M. R. Matthews, C. E.
Wieman, and E. A. Cornell, Science { 269}, 198 (1995)

K. B. Davis, M.-O. Mewes, M. R. Andrews, N. J. van
Druten, D. S. Durfee, D. M. Kurn, and W. Ketterle,
Phys. Rev. Lett. { 75}, 3969 (1995)

\bibitem{exptspinor}
J. Stenger, S. Inouye, D. M. Stamper-Kurn, H.-J. Miesner, A. P. Chikkatur, and  W. Ketterle, Nature  { 396}, 345 (1998)  


\bibitem{thso}
J. Dalibard, F. Gerbier, G. Juzeli\=unas, P. \"Ohberg, Rev. Mod. Phys. { 83}, 1523 (2011)
 


\bibitem{SOdre}
G. Dresselhaus, Phys. Rev. { 100},  580 (1955)


\bibitem{SOras}E. I. Rashba, Fiz. Tverd. Tela { 2}, 1224 (1960) [English
Transla.: Sov. Phys. Solid State { 2}, 1109 (1960)]


\bibitem{exptso}
Y.-J. Lin, K. Jim\'enez-Garc\'ia, I. B. Spielman, Nature { 471}, 83 (2011)


\bibitem{na-solid}J. Li, W. Huang, B. Shteynas, S. Burchesky, F. C. Top, E. Su, J. Lee, A. O. Jamison, and W. Ketterle,
Phys. Rev. Lett. { 117}, 185301 (2016)
  



\bibitem{exptsp1}
D. Campbell, R. Price, A. Putra, A. Vald\'es-Curiel,
D. Trypogeorgos, and I. B. Spielman,  Nat. Commun.
{ 7}, 10897 (2016)


\bibitem{sherman}I. V. Tokatly and E. Ya. Sherman
Phys. Rev. A 87, 041602(R) (2013)




\bibitem{thspinorb}
%
V. I. Yukalov, Laser Phys. { 28},  053001 (2018)

Y. Kawaguchi and M. Ueda, Phys. Rep. { 520},  253 (2012)

S. Gautam and  S. K. Adhikari, Phys. Rev. A { 92},  023616 (2015)



\bibitem{sprsld}
M. Boninsegni and N. V.  Prokof'ev,  Rev. Mod. Phys. { 84}, 759 (2012)

A. F. Andreev and I. M. Lifshitz, Zh. Eksp. Teor. Fiz. { 56}, 2057 (1969)
[English Transla.: Sov. Phys. JETP { 29},
1107 (1969)] 

E.P. Gross, Phys. Rev. {  106}, 161 (1957)

A. J. Leggett, Phys. Rev. Lett. { 25}, 1543 (1970)

 G. V. Chester, Phys. Rev. A { 2}, 256 (1970)

 V. I. Yukalov, Physics { 2}, 49 (2020)

\bibitem{fermi}M. W. Zwierlein, C. H. Schunck, A. Schirotzek, and  W. Ketterle, 
Nature  { 442}, 54 (2006)

M. Greiner, C. A. Regal, and D. S.  Jin,   Nature  { 426}, 537 (2003)

  \bibitem{solid-1/2}T.-L. Ho and S. Zhang, Phys. Rev. Lett. { 107}, 150403
(2011)  

R. Liao,
Phys. Rev. Lett. { 120}, 140403 (2018)

W. Han, X.-F. Zhang, D.-S. Wang, H.-F. Jiang, W. Zhang, and S.-G. Zhang,
Phys. Rev. Lett. { 121}, 030404 (2018)


 


\bibitem{st2}J.-R. Li, J. Lee, W. Huang, S. Burchesky, B. Shteynas, F. \c C. Top,
A. O. Jamison, and W. Ketterle, Nature  { 543}, 91 (2017)

 
\bibitem{losh}
R. Bombin, J. Boronat, and F. Mazzanti,
Phys. Rev. Lett. { 119}, 250402 (2017)
 
Zhen-Kai Lu, Yun Li, D. S. Petrov, and   G.  V.  Shlyapnikov,
Phys. Rev. Lett. { 115}, 075303 (2015)

N. Y. Yao, C. R. Laumann, A. V. Gorshkov, S. D. Bennett, E. Demler, P. Zoller, and M. D. Lukin,
Phys. Rev. Lett. { 109}, 266804 (2012) 

 



\bibitem{finite}  G. Masella, A. Angelone, F. Mezzacapo, G. Pupillo, and N. V. Prokof'ev,
Phys. Rev. Lett. { 123}, 045301 (2019)

F. Cinti, P. Jain, M. Boninsegni, A. Micheli, P. Zoller,
and G. Pupillo, Phys. Rev. Lett. { 105}, 135301 (2010)
  
S. Saccani, S. Moroni, and M. Boninsegni, Phys. Rev.
Lett. { 108}, 175301 (2012)

{
\bibitem{dipolar2d}   F. B\"ottcher, J.-N. Schmidt, M. Wenzel, J. Hertkorn, M.
Guo, T. Langen, T. Pfau, Phys. Rev. X { 9},  011051 (2019)

J. Hertkorn, F. B\"ottcher, M. Guo, J. N. Schmidt, T. Langen, H. P. B\"uchler, T. Pfau, Phys. Rev. Lett. { 123},
 193002 (2019)


    

\bibitem{dipolar1d}L. Tanzi, E. Lucioni, F. Fam\'a, J. Catani, A. Fioretti, C.
Gabbanini, R. N. Bisset, L. Santos, G. Modugno, Phys.
Rev. Lett. { 122},  130405 (2019)

 S. V. Andreev,
Phys. Rev. B { 95}, 184519  (2017)

G. Natale, R. M. W. van Bijnen, A. Patscheider, D. Petter, M. J. Mark, L. Chomaz, F. Ferlaino, Phys. Rev. Lett.
{ 123}, 050402 (2019)
}



\bibitem{kita}T. Mizushima, K. Machida, T. Kita, Phys. Rev. Lett. { 89},  030401 (2002)

T. Mizushima, K. Machida, T. Kita, Phys. Rev. A { 66},  053610 (2002)



\bibitem{2020} A. Putra, F. Salces-Carcoba, Y. Yue, S. Sugawa, I. B. Spielman, 
Phys. Rev. Lett. { 124},  053605 (2020)


\bibitem{stringari}Y. Li, G. I. Martone, L. P. Pitaevskii, and S. Stringari,
Phys. Rev. Lett. { 110}, 235302 (2013)

G. I. Martone, Y. Li, and S. Stringari, Phys. Rev. A { 90}, 041604(R) (2014)




\bibitem{adhikari}S. K. Adhikari, Phys. Lett. A { 388},  127042 (2021)

S. K. Adhikari,
Phys. Rev. A  103, L011301 (2021)

\bibitem{ad2}S. K. Adhikari, Physica E 118,  113892  (2020)
 
 


 






\bibitem{zhai}H. Zhai, Int. J. of Mod. Phys. B { 26}, 1230001 (2012)
 

\bibitem{sala}E. P. Gross,  Nuovo Cimento {\bf 20},  454 (1961);
L. P. Pitaevskii, Zurn. Eksp. Teor. Fiz. {\bf 40},  646 (1961) [English Transla.:
Sov. Phys. JETP {\bf 13},  451 (1961).]

 
\bibitem{bec2009}R. Ravisankar, D. Vudragovi\'c, P. Muruganandam, A.
Bala\v{z}, and S. K. Adhikari,  Comput. Phys. Commun. { 259}, 107657  (2020)

P. Muruganandam and S.~K. Adhikari, Comput. Phys. Commun. { 180}, 1888 (2009)

D. Vudragovi\'c, I. Vidanovi\'c, A. Bala\v{z}, P. Muruganandam, S. K. Adhikari, Comput. Phys. Commun. 
{ 183},  2021 (2012)





\bibitem{kokk} E. G. M. van Kempen, S. J. J. M. F. Kokkelmans, D. J.
Heinzen, and B. J. Verhaar, Phys. Rev. Lett. { 88}, 093201
(2002)



 \bibitem{crube}A. Crubellier,  O. Dulieu, F. Masnou-Seeuws, M. Elbs,
H. Knockel, and E. Tiemann, Eur. Phys.
J. D { 6}, 211 (1999)


\bibitem{quasi-1d1}S. Gautam and S. K. Adhikari, Phys. Rev. A { 91}, 063617 (2015)

S. Gautam and S. K. Adhikari, Laser Phys. Lett. { 12}, 045501 (2015)



\bibitem{mal}H. Sakaguchi, E. Ya. Sherman, and B. A. Malomed, Phys. Rev. E 94, 032202 (2016)

 



 


   



 
\end{thebibliography}
\end{document}